\documentclass[aps,prd,preprint,superscriptaddress,amsmath,amssymb,showpacs]{revtex4-1}
\usepackage{dcolumn}
\usepackage{graphicx}
\usepackage{float}
\usepackage{physics}
\usepackage[colorlinks=true,allcolors=blue]{hyperref}

\begin{document}

\title{Gravitational waves from holographic QCD phase transition with gluon condensate}
\author{Zhou-Run Zhu}
\email{zhuzhourun@mails.ccnu.edu.cn}
\affiliation{Institute of Particle Physics and Key Laboratory of Quark and Lepton Physics (MOS), Central China Normal University,
Wuhan 430079, China}

\author{Jun Chen}
\email{2021012@hbmzu.edu.cn}
\affiliation{Institute of Particle Physics and Key Laboratory of Quark and Lepton Physics (MOS), Central China Normal University,
Wuhan 430079, China}

\author{Defu Hou }
\thanks{Corresponding author}
\email{houdf@mail.ccnu.edu.cn}
\affiliation{Institute of Particle Physics and Key Laboratory of Quark and Lepton Physics (MOS), Central China Normal University,
Wuhan 430079, China}

\date{\today}

\begin{abstract}
In this paper, we discuss the holographic first order QCD phase transition with gluon condensate and the generation of gravitational waves (GWs) from the phase transition. The first order QCD phase transition is dual to the first order Hawking-Page phase transition from holography. We study the first order Hawking-Page phase transition from the thermal dilatonic phase to the dilatonic black hole phase and find the phase transition temperature is proportional to the gluon condensate. After substituting into the phenomenological value of gluon condensate from QCD sum rules, we find $T_c=155.38\ MeV$. In further research, we study the GWs generated from holographic cosmic first order QCD phase transition with gluon condensate and the produced GWs might be detected by the International Pulsar Timing Array, Square Kilometre Array and Big-Bang Observer. Moreover, the gluon condensate suppresses the energy density of total GWs and peak frequency.
\end{abstract}

\maketitle

\section{Introduction}\label{sec:01_intro}

From the prediction of Albert Einstein based on the general relativity, gravitational waves (GWs) serves as the ripples of space-time metric \cite{Einstein:1916cc,Einstein:1918btx}. After about one hundred years, the LIGO Scientific and Virgo Collaborations announced the GWs signal was detected directly \cite{Abbott:2016blz} from a binary black hole merger. Various astrophysical processes (Binary Systems etc.), the cosmic strings, the inflation and cosmological phase transitions (PTs) in the early universe can imprint unique signatures in GWs information \cite{Cai:2017cbj}. These sources contribute to understand the unknown aspects of the evolution of the early Universe. The early Universe may have experienced several PTs in the evolution process and GWs can be generated from a first order phase transition. During the thermal processes of the first order PT, the bubbles nucleate in a sea of metastable phase. Then the bubbles expand and ultimately collide with each other. This violent and inhomogeneous process leads to a significant stochastic background of GWs \cite{Kosowsky:1992rz,Kosowsky:1992vn,Kamionkowski:1993fg,Huber:2008hg,Caprini:2015zlo,Jinno:2015doa}.
In addition to the contribution of the bubbles collision, sound waves \cite{Hindmarsh:2013xza,Hindmarsh:2015qta} and Magnetohydrodynamic(MHD) turbulence in the plasma \cite{Kosowsky:2001xp,Caprini:2006jb,Caprini:2009yp} conduce to the production of GWs.

In order to search the possibility of investigating phase transitions, the extreme conditions of high temperature and energy density need to be created. This task can be realized by the heavy ion
collision experiments at RHIC and LHC \cite{Arsene:2004fa,Adcox:2004mh,Back:2004je,Adams:2005dq,Aad:2013xma}. During the heavy ion collision, the confinement-deconfinement QCD phase transition happens and the strongly coupled plasma forms. The results of lattice QCD show that PT is first order for heavy, static quarks or pure gauge theory \cite{Lucini:2012wq} while PT is crossover for physical quark masses \cite{Aoki:2006br,Bazavov:2011nk,Bhattacharya:2014ara}. The cosmological first order QCD PT may happen during the evolution of the early Universe and may lead to the production of GWs. Studying the production of GWs from the first order QCD PT may give some inspiration for exploring the QCD phase structure.

In this paper, we want to study the GWs generated from confinement-deconfinement PT by using the AdS/CFT correspondence \cite{Witten:1998qj,Gubser:1998bc,Maldacena:1997re}. The first order confinement-deconfinement PT corresponds to the first order Hawking-Page PT has been discussed in \cite{Hawking:1982dh}. The Schwarzschild black hole is dual to the deconfinement at high temperature while thermal AdS is dual to confinement at low temperature \cite{Witten:1998zw}. In \cite{Herzog:2006ra}, the author researches the deconfinement PT in the hard and soft wall AdS/QCD models. Based on \cite{Herzog:2006ra}, the GWs generated from cosmological QCD phase transition within hard and soft wall AdS/QCD has been studied in \cite{Ahmadvand:2017xrw}. Moreover, the effects of chemical potential and finite coupling on the production of GWs have been discussed in \cite{Ahmadvand:2017tue,Rezapour:2020mvi} from holography respectively. The GWs generated from QCD phase transition and electroweak phase transition within holographic models has been investigated in \cite{Chen:2017cyc}. The gravitational waves produced from the holographic QCD with different chemical potential has been discussed in \cite{Li:2018oqf} and GWs generated from a holographic model of five dimensional Einstein gravity which is coupled to a scalar field can be seen in \cite{Ares:2020lbt}.

In this work, we discuss the first order phase transition in the gravity-dilaton background and study the gravitational waves of holographic cosmological QCD phase transition with gluon condensate. The gluon condensate was considered as an order parameter for (de)confinement  \cite{Lee:1989qj,DElia:2002hkf,Miller:2006hr}.  To be specific, we study the cosmological first order phase transition from the thermal dilatonic $AdS$ background \cite{Csaki:2006ji} (confined phase) to the dilatonic black hole \cite{Kim:2007qk} (deconfined phase) in this work. The dilaton in the 5D gravity action is dual to gluon condensate operator $\langle G^2_{\mu\nu}\rangle$. Then we discuss the relation between phase transition temperature and gluon condensate. When using an approximately renorminvariant (renormalization group invariant) quantity $\langle \frac{\alpha_s}{\pi}G^2_{\mu\nu} \rangle$ instead of scale-dependent $\langle G^2_{\mu\nu}\rangle$, we find phase transition temperature $T_c$ is proportional to the $\langle \frac{\alpha_s}{\pi}G^2_{\mu\nu} \rangle$. It should be mentioned that \cite{Afonin:2020crk} has studied the phase transition between the thermal dilatonic background and the $AdS_5$ black hole background (gluon condensate vanishes when T $\neq$ 0). The lattice data shows that gluon condensate is non-zero at high temperature \cite{Miller:2006hr}. From this point of view, studying the phase transition from the thermal dilatonic $AdS_5$ background to the dilatonic black hole may be more precise.

In further research, we study the GWs generated from holographic cosmological first order QCD phase transition with gluon condensate. From the results, the produced total GWs might be detected by the International Pulsar Timing Array (IPTA), Square Kilometre Array (SKA) and Big-Bang Observer (BBO). Moreover, the contribution of the bubble collisions to the total GWs spectrum is dominant in the $f< 4\times 10^{-8}$ Hz and $f> 1.5\times 10^{-6}$ Hz region. The contribution of sound waves to the total GWs spectrum is around the region of $ 4\times 10^{-8}$ Hz $< f$ $<1.5\times 10^{-6}$ Hz. The peak frequency is determined by the sound waves and located around $2\times 10^{-7}$ Hz. The contribution of MHD turbulence to the total GWs spectrum in our results is negligible. Moreover, the gluon condensate suppresses the peak value of total GWs frequency spectrum and peak frequency.

The paper is organized as follows. In Sec.~\ref{sec:02}, we study the holographic first order QCD phase transition with gluon condensate. In Sec.~\ref{sec:03}, we investigate the gravitational waves of a first order phase transition with gluon condensate. The conclusion and discussion are given in Sec.~\ref{sec:04}.

\section{Holographic QCD phase transition with gluon condensate}\label{sec:02}

In this section, we discuss de/confinement (first order) phase transition with gluon condensate. More specifically, we investigate the Hawking-Page phase transition from the thermal dilatonic $AdS$ background (confined phase) to the dilatonic black hole (deconfined phase). First, we consider the back-reaction of the metric to the gluon condensate. The 5D gravity action with a dilaton field is
\begin{equation}\label{eq:x1}
S=-\frac{1}{2k^2} \int d^5x\sqrt{g} \bigg(\mathcal{R}+\frac{12}{L^2}-\frac{1}{2}\partial _\mu \phi \partial ^\mu \phi\bigg),
\end{equation}
where dilaton $\phi$ is coupled to the gluon operator. $\mathcal{R}$ denotes the Ricci scalar and $L$ is the AdS radius.

There are two relevant solutions of this gravity action. One is the dilaton-wall solution at zero temperature. In this thermal dilatonic AdS (tdAdS) background \cite{Csaki:2006ji}, the metric is
\begin{align}
\label{eq:x2}
  ds^2&=\frac{L^2}{z^2}\bigg(\sqrt{1-c^2z^8}(d\vec{x}^2+dt^2)+dz^2 \bigg),
\end{align}
and the dilaton is given by
\begin{align}
\label{eq:x3}
\phi(z)&=\sqrt{\frac{3}{2}}log\bigg(\frac{1+cz^4}{1-cz^4}\bigg)+\phi_0,
\end{align}
where $\phi_0$ is just a constant. $c$ represents the gluon condensate and relates the the gluon operator $G^2_{\mu\nu}$
\begin{equation}\label{eqv}
\langle G^2_{\mu\nu} \rangle=\frac{8\sqrt{3(N^2_c -1)}}{\pi} c.
\end{equation}

From Eq.(\ref{eq:x3}), one can find there is a singularity in the dilaton. The singularity is at $z_c =c^{-\frac{1}{4}}$. Therefore the solution is well defined in the range $0< z <z_c$. $z_c$ can be seen as the IR cut-off of the tdAdS background.

Next we evaluate the gravity and dilaton action as
\begin{equation}\label{eqh}
\mathcal{R}+\frac{12}{L^2}=-\frac{8}{L^2}+\frac{48c^2 z^8}{L^2 (1-c^2 z^8)^2},\\
-\frac{1}{2}\partial _\mu \phi \partial ^\mu \phi=-\frac{48c^2 z^8}{L^2 (1-c^2 z^8)^2}.
\end{equation}

From above calculation, one can find the singularity of the dilaton part is counteracted by the singularity generated in the back-reaction to the gluon condensation of gravity part. Namely, the total action (Eq.(\ref{eq:x1})) has no singularity for any finite values of $c$. The IR divergence of the dilaton has no influence on the existence of gluon condensate.

Then one can calculate the action density of the thermal dilatonic AdS background (tdAdS)
\begin{equation}\label{eqi}
I_1 = -\frac{1}{2k^2} \int^{\beta'}_0 dt \int^{z_c}_\epsilon \frac{L^5}{z^5}(-\frac{8}{L^2})(1-c^2 z^8)dz.
\end{equation}

The other solution of the dilatonic black hole (dBH) at finite tempeature $T$ is given by \cite{Kim:2007qk}
\begin{equation}\label{eq:x4}
 ds^2=\frac{L^2}{z^2}\bigg((1-f^2 z^8)^{\frac{1}{2}}(\frac{1+f z^4}{1-f z^4})^{\frac{a}{2f}} (d\vec{x}^2- (\frac{1-f z^4}{1+f z^4})^{\frac{2a}{f}}  dt^2)+dz^2\bigg),
\end{equation}
where
\begin{align}\label{eq:x5}
f^2=a^2+c^2,\ a=\frac{1}{4}(\pi T)^4.
\end{align}

In gravity theory, the temperature $T$ determine the scale of the metric. As discussed in \cite{Kim:2007qk}, $T$ is the first non-trivial coefficient in the metric expansion $g^{(4)}_{\mu\nu}$. $g^{(4)}_{\mu\nu}$ is the metric expansion with power of $z^4$ and does not change when $c$ is non-zero from the results.

\begin{align}\label{eq:x51}
g_{00} = -1+3az^4+ o(z^8),\ g_{ii}=1+az^4+o(z^{12}).
\end{align}

It could be explained as: Gluons are massless excitation when participating in thermal excitations. In this process, the gluons follow the Stephan-Boltzmann law $\rho\sim T^4$. Putting aside the gluon condensation contribution, it is reasonable to identify $g^{(4)}_{\mu\nu}$ as the gluon contribution to the energy momentum tensor $T_{\mu\nu}$.

\begin{align}\label{eq:x52}
g^{(4)}_{\mu\nu} = a(3,1,1,1)= \frac{\kappa^2}{2}diag(\rho,g,g,g)_{gluon}.
\end{align}

This supports $a=\frac{1}{4}(\pi T)^4$ since the thermal gluons follow the Stephan-Boltzmann law.

The corresponding dilaton profile is
\begin{equation}\label{eq:x6}
  \phi(z)=\frac{c}{f}\sqrt{\frac{3}{2}} log \bigg(\frac{1+fz^4}{1-fz^4}\bigg)+\phi_0,
\end{equation}
where the singularity is at $z_f =f^{-\frac{1}{4}}$. This solution is valid only in the range $0< z <z_f$. $z_f$ can be regarded as the IR cut-off of the dBH.

The gravity and dilaton parts are
\begin{equation}\label{eqh1}
\mathcal{R}+\frac{12}{L^2}=-\frac{8}{L^2}+\frac{48c^2 z^8}{L^2 (1-f^2 z^8)^2},\\
-\frac{1}{2}\partial _\mu \phi \partial ^\mu \phi=-\frac{48c^2 z^8}{L^2 (1-f^2 z^8)^2}.
\end{equation}

In dBH, the singularity is also canceled out by the dilaton part and gravity part.

The action density of the dilatonic black hole (dBH) is
\begin{equation}\label{eqg}
I_2 = -\frac{1}{2k^2} \int^{\beta}_0 dt \int^{z_f}_\epsilon \frac{L^5}{z^5}(-\frac{8}{L^2})(1-f^2 z^8)dz.
\end{equation}

Using the period relations of the two backgrounds: $\beta'\simeq \beta (1-\frac{3}{2}a \epsilon^4)$ \cite{Kim:2007qk}, we can calculate the free energy difference between tdAdS phase and dBH phase in the limit $\epsilon =0$

\begin{equation}\label{eqb}
\Delta F = \frac{L^3}{2k^2}(3a+4c-4f),
\end{equation}
where $L^3 / k^2 = N^2_c/4\pi^2$ \cite{Sin:2007ze}.

Hawking-Page transition happens at $\Delta F = 0$. Different from \cite{Kim:2007qk}, we want to study the relations between the phase transition temperature and gluon condensate. In renormalization group approach, it is necessary to present the relevant quantities in a renorminvariant (renormalization group invariant) form to avoid the quantities are scale-dependent \cite{Nesterenko:2000veu}. Since $\langle G^2_{\mu\nu} \rangle$ is scale-dependent, we use the approximately renorminvariant quantity $\langle \frac{\alpha_s}{\pi}G^2_{\mu\nu} \rangle$ (the one-loop approximation to renorminvariant $\langle \beta G^2_{\mu\nu} \rangle$ and $\beta$ = $\beta(\alpha_s)$ represents the QCD $\beta$-function).From Eq.(\ref{eqv}) and (\ref{eqb}), one can get $T^4_c$ is in proportion to $\langle \frac{\alpha_s}{\pi}G^2_{\mu\nu} \rangle$ when we set $N_c =3$

\begin{equation}\label{eqx}
T^4_c = \frac{\sqrt{6}}{7}\frac{\langle \frac{\alpha_s}{\pi}G^2_{\mu\nu} \rangle}{\pi^2 \alpha_s},
\end{equation}
where $\alpha_s$ is gauge coupling.

In order to fix $\alpha_s$, one could take the values of $T_c$ and gluon condensate $\langle \frac{\alpha_s}{\pi}G^2_{\mu\nu} \rangle$ in pure $SU(3)$ gluodynamics from lattice simulations.
The relevant lattice results for $SU(3)$ Yang-Mills theory are $T_c=264\ MeV$ \cite{Boyd:1996bx} and $\langle \frac{\alpha_s}{\pi}G^2_{\mu\nu} \rangle = 0.1\ GeV^4$ \cite{Campostrini:1989uj}. Then one can get $\alpha_s=0.73$. Next, we want to discuss $T_c$ in the real world with physical quarks. The phenomenological estimation of the gluon condensate from QCD sum rules is $\langle \frac{\alpha_s}{\pi}G^2_{\mu\nu} \rangle = 0.012\ GeV^4$ \cite{Shifman:1978bx}. Using this estimation, one can get $T_c=155.38\ MeV$. It should be mentioned that the authors do not predict the values of $c$ in \cite{Kim:2007qk} while we get $T_c=155.38\ MeV$ by using the phenomenological estimation of the gluon condensate from QCD sum rules. From the experimental point of view, the chemical freeze-out temperature $T_{cf}$ is considered to be close to the deconfinement temperature $T_{c}$. The deconfinement temperature of lattice simulations is $154\pm 9 \ MeV$ \cite{Bazavov:2011nk,Bazavov:2014pvz}. $T_{cf}$ extracted from the results of relativistic nuclear collisions is $156.5\pm 1.5 \ MeV$ \cite{Andronic:2017pug}. Our result is consistent with the lattice simulations results and experimental results. The five-dimensional gravitational constant $\kappa$ and five-dimensional coupling constant $g$ relate to the number of color $N_c$ and flavor $N_f$ as $L^3 / \kappa^2 = N^2_c/4\pi^2$ and $L / g^2 = N_c N_f /4\pi^2$. We find only $\kappa$ exists in this gravitational action. The action of Eq.(1) is dual to pure gluodynamics which means the $N_f$ is zero. The gauge coupling $\alpha_s$ in Eq.(15) is related to the QCD $\beta$-function. If determining $\alpha_s$ from $\beta$-function, the value of $\alpha_s$ is parameters (momentum, $\Lambda_{QCD}$) dependent. In order to get the value of $\alpha_s$ more precisely, we determine $\alpha_s$ from the results of lattice data on $T_c$ and lattice simulations on gluon condensate directly. We want to discuss $T_c$ in the real world with physical quarks and input the phenomenological estimation of the gluon condensate from QCD sum rules. From the point view of bottom-up holographic approach, $N_f$ can be considered as three in this case and the obtained $T_c$ is consistent with lattice and experimental results.

In this section, we studied the Hawking-Page phase transition from the thermal dilatonic background to the dilatonic black hole. We calculated the free energy difference between thermal dilatonic phase and dilatonic black hole phase. Then we used the phenomenological estimation of the gluon condensate from QCD sum rules and get $T_c=155.38\ MeV$ which agreed with the lattice simulations and experimental results. Note that the Hawking-Page phase transition between the thermal dilatonic background and the $AdS_5$ black hole background (gluon condensate vanishes when T $\neq$ 0) has been studied in \cite{Afonin:2020crk}. The deconfinement temperature $T_c$ is $156\ MeV$ in \cite{Afonin:2020crk}. However, the lattice data showed that gluon condensate is non-zero at high temperature \cite{Miller:2006hr}. From this point of view, studying the Hawking-Page phase transition from the thermal dilatonic background to the dilatonic black hole may be more precise.

\section{Gravitational waves from holographic QCD phase transition}\label{sec:03}

In this section, we want to study the gravitational waves from holographic cosmological first-order QCD phase transition with gluon condensate. Gravitational waves (GWs) can be produced from the first order phase transition (PT). When the first order PT happens in a thermal bath, bubbles nucleate. Then the bubbles expand in the plasma since the vacuum energy released from the initial phase and collide with each other. During the bubbles collision, spherical symmetry of the bubbles is broken and part of the stored energy leads to the GWs generation. Indeed, sound waves and Magnetohydrodynamic (MHD) turbulence in the plasma after the bubbles collide contribute to the production of GWs. The contribution of bubbles collision ($h^{2}\Omega_{env}$), sound waves ($h^{2}\Omega_{sw}$) and MHD turbulence ($h^{2}\Omega_{turb}$) to the total GWs frequency spectrum ($h^{2}\Omega(f)$) can be given by \cite{Caprini:2015zlo}

\begin{equation}\label{a}
h^{2}\Omega(f)=h^{2}\Omega_{env}(f)+h^{2}\Omega_{sw}(f)+h^{2}\Omega_{turb}(f)~,
\end{equation}
where
\begin{equation}\label{b}
\begin{split}
 &h^{2}\Omega_{env}(f)=1.67\times10^{-5}\left(\frac{0.11v_{b}^{3}}{0.42+v_{b}^{2}}\right)\left(\frac{H_{\ast}}{\tau}\right)^{2}\left(\frac{\kappa\alpha}{1+\alpha}\right)^{2}\left(\frac{100}{g_{\ast}}\right)^{\frac{1}{3}}S_{env}(f),\\
&h^{2}\Omega_{sw}(f)=2.65\times10^{-6}\left(\frac{H_{\ast}}{\tau}\right)\left(\frac{\kappa_{v}\alpha}{1+\alpha}\right)^{2}\left(\frac{100}{g_{\ast}}\right)^{\frac{1}{3}}v_{b}S_{sw}(f),\\
&h^{2}\Omega_{turb}(f)=3.35\times10^{-4}\left(\frac{H_{\ast}}{\tau}\right)\left(\frac{\kappa_{turb}\alpha}{1+\alpha}\right)^{\frac{3}{2}}\left(\frac{100}{g_{\ast}}\right)^{\frac{1}{3}}v_{b}S_{tu}(f).
\end{split}
\end{equation}

The spectral shapes of GWs are given by
\begin{equation}\label{c}
\begin{split}
&S_{env}(f)=\frac{3.8\left(\frac{f}{f_{env}}\right)^{2.8}}{1+2.8\left(\frac{f}{f_{env}}\right)^{3.8}},
\\
&S_{sw}(f)=\left(\frac{f}{f_{sw}}\right)^{3}\left(\frac{7}{4+3\left(\frac{f}{f_{sw}}\right)^{2}}\right)^{\frac{7}{2}},
\\
&S_{turb}(f)=\frac{\left(\frac{f}{f_{turb}}\right)^{3}}{\left(1+\frac{f}{f_{turb}}\right)^{\frac{11}{3}}\left(1+\frac{8\pi f}{h_{\ast}}\right)},
\end{split}
\end{equation}
with
\begin{equation}\label{d}
h_{\ast}=16.5\times10^{-6}[\text{Hz}]\left(\frac{T_{\ast}}{100\text{GeV}}\right)\left(\frac{g_{\ast}}{100}\right)^{\frac{1}{6}}.
\end{equation}

The peak frequency of each GWs spectrum is given by
\begin{equation}\label{e}
\begin{split}
&f_{env}=16.5\times10^{-6}[\text{Hz}]\left(\frac{f_{\ast}}{\tau}\right)\left(\frac{\tau}{H_{\ast}}\right)\left(\frac{T_{\ast}}{100\text{GeV}}\right)\left(\frac{g_{\ast}}{100}\right)^{\frac{1}{6}},
\\
&f_{sw}=1.9\times10^{-5}[\text{Hz}]\left(\frac{1}{v_{b}}\right)\left(\frac{\tau}{H_{\ast}}\right)\left(\frac{T_{\ast}}{100\text{GeV}}\right)\left(\frac{g_{\ast}}{100}\right)^{\frac{1}{6}},
\\
&f_{turb}=2.7\times10^{-5}[\text{Hz}]\left(\frac{1}{v_{b}}\right)\left(\frac{\tau}{H_{\ast}}\right)\left(\frac{T_{\ast}}{100\text{GeV}}\right)\left(\frac{g_{\ast}}{100}\right)^{\frac{1}{6}},
\end{split}
\end{equation}
where
\begin{equation}\label{f}
\frac{f_{\ast}}{\tau}=\frac{0.62}{1.8-0.1v_{b}+v_{b}^{2}}.
\end{equation}

In Eq.(\ref{b}), $\kappa$ denotes the fraction of the vacuum energy transformed into the kinetic energy of the bubbles. $\kappa_v$ represents the fraction which converted into bulk fluid motion. $\kappa_{tu}$ is the fraction of the latent heat converted to MHD turbulence. The expressions are given by
\begin{equation}\label{i}
\kappa=1-\frac{\alpha_\infty}{\alpha},~\kappa_{v}=\frac{\alpha_\infty}{\alpha}\frac{\alpha_\infty}{0.73+0.083\sqrt{\alpha_\infty}+\alpha_\infty},~\kappa_{tu}=\varepsilon \kappa_{v},
\end{equation}
where $\varepsilon$ denotes the fraction of bulk motion which is turbulent. Recent results suggest that $\varepsilon$ is about 0.05 $-$ 0.1 \cite{Hindmarsh:2015qta} and we take it 0.05 in the calculations.

$\alpha$ is the ratio of vacuum energy density to the thermal energy density and $\alpha_\infty$
is the minimum value of $\alpha$ that bubbles can run away
\begin{equation}\label{h}
\alpha=\frac{\epsilon_{\ast}}{\frac{\pi^{2}}{30}g_{\ast}T_{\ast}^{4}},\ \alpha_\infty= \frac{30}{24\pi^2} \frac{\Sigma_a c_a \Delta m^2_a}{g_{\ast}T_{\ast}^{2}}.
\end{equation}
where $T_{\ast}$ is the first order PT temperature. Follow \cite{Ahmadvand:2017xrw,Ahmadvand:2017tue}, we also assume $T_{\ast} \simeq T_{c}$. Moreover, $g_{\ast}$ is the number of effective relativistic degrees of freedom and $g_{\ast}\sim 10$ at the PT. $c_{a}= N_{a}$ for bosons and $c_{a}=\frac{1}{2} N_{a}$ for fermions. The $N_{a}$ is the number of the degrees of freedom for the species $a$. $\Delta m_a$ represents the mass difference of the particles between two phases and we take $\Delta m_a \sim 400\ MeV$ \cite{Lavelle:1995ty,BorkaJovanovic:2010yc}.

$\epsilon_{\ast}$ is the related vacuum energy (latent heat)
\begin{equation}\label{i1}
\epsilon_{\ast}= \bigg(-\Delta F(T)+T \frac{d \Delta F(T)}{dT}\bigg)\bigg|_{T=T_{\ast}},
\end{equation}
where $\Delta F$ is the free energy difference of two phases.

$\tau^{-1}$ denotes the duration of the PT. The hubble parameter at the temperature $T_{\ast}$ is
\begin{equation}\label{j}
H_{\ast}= \sqrt{\frac{8\pi^3 g_{\ast}}{90}} \frac{T^2_{\ast}}{m_p}.
\end{equation}
where $m_p$ is Planck mass.

\begin{figure}[H]
    \centering
      \setlength{\abovecaptionskip}{0.1cm}
    \includegraphics[width=10cm]{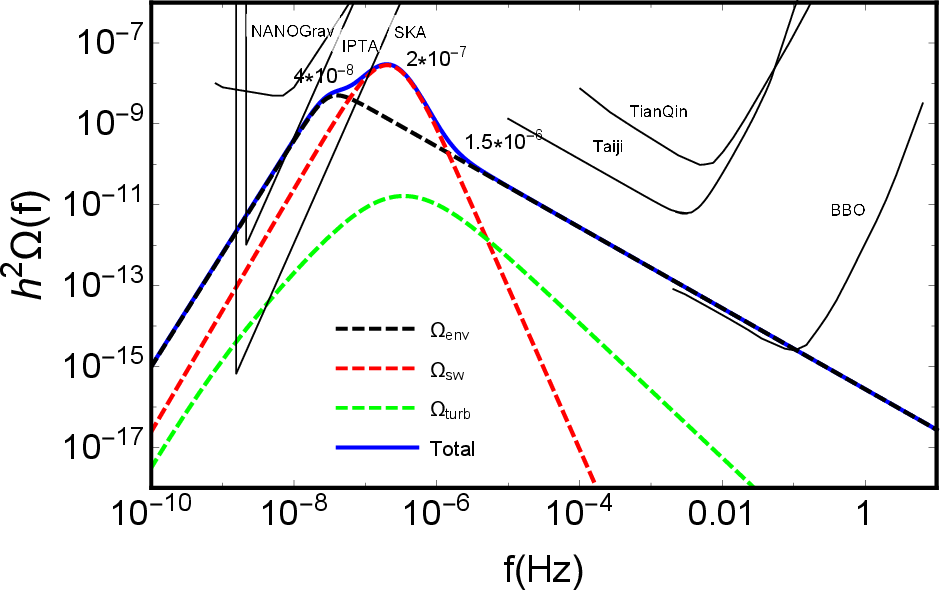}
    \caption{\label{fig1}  The GWs spectrum produced from the first order QCD phase transition with gluon condensate. The contribution of bubble collision (dashed black line), sound waves (dashed red line) and MHD turbulence (dashed green line) are plotted in this figure respectively. The solid blue line denotes the total frequency spectrum.}
\end{figure}

For $\alpha < \alpha_\infty$, the contribution of bubble collision is negligible. In $\alpha >\alpha_\infty$ case, bubbles runaway and the bubble wall velocity can reach to the speed of light. In this situation, all three sources contribute to the GWs frequency spectrum when $\alpha >\alpha_\infty$, thus $h^{2}\Omega(f)=h^{2}\Omega_{env}(f)+h^{2}\Omega_{sw}(f)+h^{2}\Omega_{turb}(f)$. Using Eq.(\ref{eqb},\ref{h},\ref{i1}), one can calculate the values of $\alpha$ and $\alpha_\infty$. We find $\alpha >\alpha_\infty$ in the presence of gluon condensate when set $\langle \frac{\alpha_s}{\pi}G^2_{\mu\nu} \rangle = 0.012\ GeV^4$. Thus the bubble wall velocity $v_b =1$ in this situation. We take typical value of $\tau/H_{\ast}=10$ in numerical calculation \cite{Ahmadvand:2017xrw}.

In Fig.~\ref{fig1}, we plot the GWs spectrum produced from the cosmological first order QCD phase transition with gluon condensate when $\langle \frac{\alpha_s}{\pi}G^2_{\mu\nu} \rangle = 0.012\ GeV^4$. The critical temperature of the first order phase transition from Eq.(\ref{eqx}) is $155.38 \ MeV$. From the results, the total GWs spectrum might be detected by the International Pulsar Timing Array (IPTA), Square Kilometre Array (SKA) and Big-Bang Observer (BBO) \cite{Moore:2014lga}. However, the total GWs spectrum might not be detected by the North American NanoHertz Observatory for Gravitational Waves (NANOGrav), TianQin and Taiji \cite{Schmitz:2020syl,Ruan:2018tsw,NANOGrav:2020bcs}. Specifically, the spectrum of bubbles collision and total GWs spectrum coincide in the frequency regions $f< 4\times 10^{-8}$ Hz and $f> 1.5\times 10^{-6}$ Hz which indicates the contribution of the bubble collisions to the total GWs spectrum is dominant in these regions. The contribution of sound waves to the total GWs is around the region of $ 4\times 10^{-8}$ Hz $< f$ $<1.5\times 10^{-6}$ Hz. We find that the peak frequency is determined by the sound waves and located around $2\times 10^{-7}$ Hz. The contribution of MHD turbulence to the total GWs spectrum is insignificant from the results.

\begin{figure}[H]
    \centering
      \setlength{\abovecaptionskip}{0.1cm}
    \includegraphics[width=10cm]{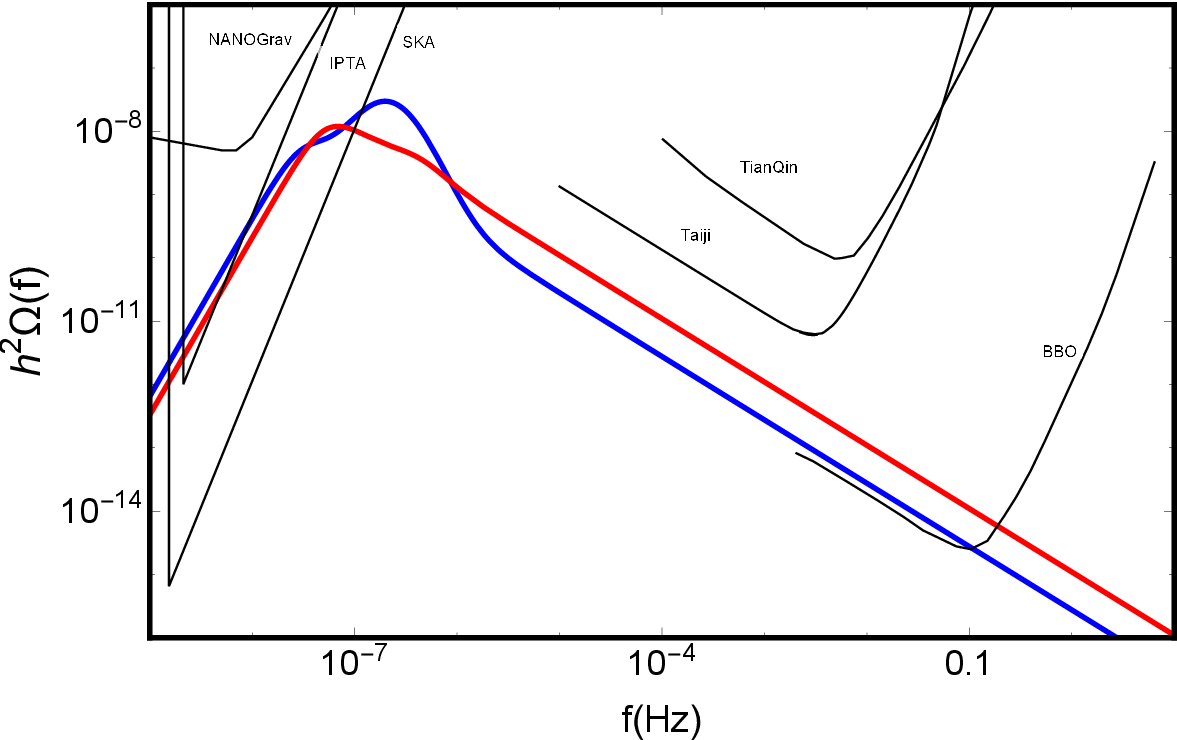}
    \caption{\label{fig2}  The total GWs frequency spectrum produced from the first order QCD phase transition with different values of gluon condensate. The blue line denotes $\langle \frac{\alpha_s}{\pi}G^2_{\mu\nu} \rangle = 0.012\ GeV^4$. The red line represents $\langle \frac{\alpha_s}{\pi}G^2_{\mu\nu} \rangle = 0.1\ GeV^4$.}
\end{figure}
In Fig.~\ref{fig2}, we plot the total GWs frequency spectrum produced from the cosmological first order QCD phase transition with different values of gluon condensate. From this figure, we find the energy density of GWs decreases as gluon condensate increases. Moreover, the position of peak frequency shifts backward when increasing the gluon condensate. It means the peak frequency decreases as gluon condensate increases. From above results, one can obtain that the gluon condensate suppresses the energy density of total GWs and peak frequency.
\section{Conclusion and discussion}\label{sec:04}

The cosmological first order QCD PT may happen during the early Universe and may lead to the production of GWs. Studying the production of GWs from the first order QCD PT may give some inspiration for exploring the QCD phase structure.

In this paper, we study the holographic cosmological first order phase transition with gluon condensate. The first order confinement-deconfinement PT corresponds to the first order Hawking-Page PT. We calculate the free energy difference between thermal dilatonic phase and the dilatonic black hole phase. We study the Hawking-Page phase transition and find phase transition temperature $T_c$ is proportional to the gluon condensate. The gauge coupling $\alpha_s$ is fixed from lattice results in SU(3) pure Yang-Mills theory. After substituting into the phenomenological value of gluon condensate in QCD sum rules, we get $T_c=155.38\ MeV$ which agrees with the lattice simulations  and experimental results.

In further research, we study the GWs generated from holographic cosmological first order QCD phase transition with gluon condensate and the produced GWs might be detected by the IPTA, SKA and BBO. Moreover, we discuss the contribution of sound waves, bubble collisions and MHD turbulence to the total GWs spectrum respectively. We find that the contribution of sound waves to the total GWs is in peak frequency region while the contribution of the bubble collisions is away from the peak frequency region. The contribution of MHD turbulence to the total GWs spectrum is negligible. Moreover, the gluon condensate suppresses the peak value of total GWs frequency spectrum and peak frequency.

\section*{Acknowledgments}

Defu Hou is in part supported by the NSFC Grant Nos. 11735007, 11890711 and 11890710.

\end{document}